\documentclass[lettersize,journal]{IEEEtran}
\usepackage{amsmath,amsfonts}
\usepackage{algorithmic}
\usepackage{algorithm}
\usepackage{array}
\usepackage[caption=false,font=normalsize,labelfont=sf,textfont=sf]{subfig}
\usepackage{textcomp}
\usepackage{stfloats}
\usepackage{url}
\usepackage{verbatim}
\usepackage{graphicx}
\usepackage{cite}
\usepackage{amssymb}

\newtheorem{defn}{Definition}

\begin{document}

\title{Physically Analyzable AI-Based Nonlinear Platoon\\Dynamics Modeling During Traffic Oscillation: \\ A Koopman Approach}

\author{$\text{Kexin Tian}^1$, $\text{Haotian Shi}^1$, $\text{Yang Zhou}^*$,~\IEEEmembership{Member,~IEEE,} and Sixu Li
\thanks{$^1$Kexin Tian and Haotian Shi contribute to the paper equally.}
\thanks{$^*$Corresponding author: Yang Zhou.}
\thanks{Haotian Shi is with the Department of Civil and Environmental Engineering, University of Wisconsin, Madison WI 53705 USA (email: hshi84@wisc.edu)}
\thanks{Kexin Tian, Yang Zhou, and Sixu Li are with the Zachry Department of Civil and Environmental Engineering, Texas A\&M University, College Station TX 77483 USA (email: ktian6@tamu.edu; yangzhou295@tamu.edu; sixuli@tamu.edu) }
}

\markboth{IEEE Transactions on Intelligent Transportation Systems}%
{Shell \MakeLowercase{\textit{et al.}}: A Sample Article Using IEEEtran.cls for IEEE Journals}

\maketitle

\begin{abstract}
Given the complexity and nonlinearity inherent in traffic dynamics within vehicular platoons, there exists a critical need for a modeling methodology with high accuracy while concurrently achieving physical analyzability. Currently, there are two predominant approaches: the physics model-based approach and the Artificial Intelligence (AI)--based approach. Knowing the facts that the physical-based model usually lacks sufficient modeling accuracy and potential function mismatches and the pure-AI-based method lacks analyzability, this paper innovatively proposes an AI-based Koopman approach to model the unknown nonlinear platoon dynamics harnessing the power of AI and simultaneously maintain physical analyzability, with a particular focus on periods of traffic oscillation. Specifically, this research first employs a deep learning framework to generate the embedding function that lifts the original space into the embedding space. Given the embedding space descriptiveness, the platoon dynamics can be expressed as a linear dynamical system founded by the Koopman theory. Based on that, the routine of linear dynamical system analysis can be conducted on the learned traffic linear dynamics in the embedding space. By that, the physical interpretability and analyzability of model-based methods with the heightened precision inherent in data-driven approaches can be synergized. Comparative experiments have been conducted with existing modeling approaches, which suggests our method's superiority in accuracy. Additionally, a phase plane analysis is performed, further evidencing our approach's effectiveness in replicating the complex dynamic patterns. Moreover, the proposed methodology is proven to feature the capability of analyzing the stability, attesting to the physical analyzability.
\end{abstract}


\section{Introduction \label{Sec.1}}
\IEEEPARstart{T}{raffic} flow dynamics, which pertains to the spatiotemporal evolution of traffic disturbances, has been the subject of extensive investigation over decades \cite{gerlough1976traffic,elefteriadou2014introduction,treiber2013traffic}. Researchers have consistently endeavored to develop increasingly sophisticated models \cite{greenberg1959analysis,khan1999modeling,hoogendoorn2013traffic,li2017vehicle,li2020trajectory}, with the aim of more accurately capturing the evolution from intricate vehicle driving behaviors. These efforts encompass a thorough analysis of individual driver characteristics \cite{sharma2015lattice,chakroborty2004microscopic} and the stochastic nature \cite{schreckenberg1995discrete,jabari2012stochastic} of traffic dynamics. Despite these considerable efforts, the nonlinear and unknown nature of driving behaviors have posed challenges in achieving a comprehensive understanding and accurate predictions of traffic flow dynamics (i.e., how the disturbance evolves over vehicular platoons)\cite{li2012microscopic,tyagi2009review,li2024sequencing}. 

The challenges over accurate modeling and prediction on disturbances evolving over platoons become more pronounced in traffic oscillations, also known as  "stop-and-go" traffic\cite{li2011characterization,mauch2002freeway,ahn2007freeway}. Traffic oscillation denotes a recurring pattern observed in congested traffic, where vehicles alternately transition between slow-moving and fast-moving phases instead of maintaining a steady state traffic flow \cite{li2010measurement,ahn2005formation}, demonstrating complex nonlinear features. To address the inherent nonlinearities within traffic flow dynamics, a substantial body of research has been devoted to the analysis of car-following behavior \cite{gazis1959car,lee1966generalization, ossen2011heterogeneity,chen2012behavioral,treiber2017intelligent,wen2023stochastic}. These analysis frameworks seek to elucidate the mechanisms by the car following nature, in which one vehicle follows another, and, by extension, aim to delineate how disturbances propagate through strings of vehicles during traffic oscillations, thereby influencing traffic flow. Research endeavors in this domain are majorly manifested through two different directions: physics model-based approaches \cite{treiber2000congested,bando1995dynamical} and artificial intelligence (AI) based approaches \cite{shi2022deep,kim2017probabilistic,huang2018car,wang2017capturing}.

Physics model-based approaches, such as the Intelligent Driver Model (IDM) \cite{treiber2000congested} and the Optimal Velocity Model (OVM) \cite{bando1995dynamical}, aim to characterize a range of car following behaviors relying on specific nonlinearity with interpretive parameters. These parameters include driver attributes such as fluctuating car-following response times, desired following distance, and sensitivity factor. Although these physics models provide a foundational structure for understanding traffic dynamics, their simplified assumptions often fail to capture the complex nonlinearity and meanwhile model mismatch may also happen. This discrepancy can limit the accuracy of these models in replicating actual traffic flow dynamics.

In AI approaches\cite{pecher2016data,papathanasopoulou2015towards}, the emphasis is on learning the traffic flow dynamics from data without explicitly specifying a mathematical model that represents the underlying physics or dynamics. This approach directly uncovers the relationships between the historical traffic flow dynamics and its subsequent evolving dynamics. Among these data-driven techniques, neural network (NN)-based methods \cite{shi2022integrated} encompass Recurrent Neural Network (RNN)-based car-following models \cite{kim2017probabilistic}. These RNN-based models include Long Short-Term Memory (LSTM)-based car-following models \cite{huang2018car,morton2016analysis} and Gated Recurrent Units (GRU)-based models \cite{wang2017capturing}. Additionally, there are Feed-forward Neural Networks (FNN)-based car-following models \cite{sarkar2017trajectory,colombaroni2014artificial}. However, these data-driven methods have their inherent limitations. While they excel at learning from empirical data and reproducing accurate car-following behaviors in the traffic flow, they often fall short in capturing the intricate nonlinear dynamics of the traffic flow in an analyzable expression which renders challenges to unveil the properties of traffic dynamics. Besides the pure AI approach, some data-driven methods such as Dynamic Mode Decomposition (DMD) \cite{kutz2016dynamic} offer an alternative approach to model traffic flow dynamics assuming the law following a linear dynamical system, which makes them accessible to a broader array of linear systems analytical tools. However, DMD inherently assumes linearity within the system they aim to characterize, which can present challenges when dealing with traffic flow dynamics' high-dimensional nonlinear nature.

To this end, there is a critical need for a comprehensive analytical tool that integrates the strengths of both data-driven and physics-based methodologies. Such a tool would be capable of unveiling the intricacies of traffic flow dynamics in a way grounded in rigorous physical analysis and enhanced by the adaptability and high accuracy of data-driven approaches. Hence, it innovatively proposes an AI-based Koopman approach to model the unknown nonlinear platoon dynamics, harnessing the power of AI and simultaneously maintaining physical analyzability, with a particular focus on periods of traffic oscillation. Specifically, this research first employs a deep learning framework to generate the embedding function that lifts the original space into the embedding space. Given the embedding space descriptiveness, the platoon dynamics can be expressed as a linear dynamical system founded by the Koopman theory\cite{koopman1932dynamical,brunton2021modern,mauroy2020koopman}. Based on that, the routine of linear dynamical system analysis can be conducted on the learned traffic linear dynamics in the embedding space. By that, the physical interpretability and analyzability of model-based methods with the heightened precision inherent in data-driven approaches can be synergized. The Koopman operator maps the state space of a dynamical system to a higher-dimensional space, allowing for effective linearization of the system's dynamics \cite{koopman1932dynamical,brunton2021modern,mauroy2020koopman}. 
The contributions of our proposed framework are summarized as follows:

\begin{enumerate}
    \item \textbf{A novel framework for modeling traffic flow dynamics in vehicular platoons}: We present an innovative framework that models the nonlinear dynamics of traffic flow, particularly in the context of traffic oscillations. This framework effectively transforms complex dynamics into a linear representation, allowing for precise reproduction and analyzable expressions of evolving traffic flow dynamics.
    
    \item \textbf{Synergy of high precision and physical interpretability}: By leveraging a deep learning framework, our approach uniquely combines the accuracy of data-driven methodologies with the critical aspects of physical interpretability and analyzability found in model-based approaches. This integration facilitates a thorough understanding of traffic flow dynamics and provides a foundation for future advancements in traffic flow prediction and control.
    
\end{enumerate}

The subsequent sections of the paper is organized as follows:  Section 2 gives the problem statement and details the proposed methodology in-depth. Section 3 describes the experiment settings. Section 4 presents the findings from our experimental evaluation.

\section{Methodology \label{Sec.2}}
\subsection{Problem Statement \label{Sec.2.1}}

Regarding traffic flow dynamics, we focus on a scenario involving a vehicular platoon situated within a single-lane highway. The objective is to model the trajectories of the following vehicles in the platoon over a designated future time span, given the initial state of these vehicles and the acceleration sequence of the leading vehicle as inputs.

We define the leading vehicle as vehicle $0$ and following vehicles as vehicle $i$, where $i \in [1, 2, \ldots, n]$ and $n \in \mathbb{R}^+$ is the total number of following vehicles. Our analysis is based on a discrete time step horizon $[0,1,\cdots,K]$. Regarding the input parameters, for each time step $k \in [0,1,\cdots,K]$ we denote a state vector $X_{k} = \{\mathbf{x}_{1, k}, \mathbf{x}_{2, k}, \ldots, \mathbf{x}_{n, k}\}$, where $\mathbf{x}_{i, k}$ denotes the state of vehicle $i \in [1, 2, \ldots, n]$ at time step $k$. Based on this, $X_0$ denotes the initial state of following vehicles in the platoon. The acceleration for the leading vehicle at time step $k \in [0,1,\cdots,K]$ is denoted as $\mathbf{u}_{k}$, serving as the control input. 

Regarding the output derived from the traffic flow modeling, we generate at the forthcoming time horizon $[1, 2, \cdots, K]$. The generated trajectories at time step $t \in [1,2, \cdots, K]$ is denoted as $\hat{Y}_{k} = \{\hat{Y}_{1,k}, \hat{Y}_{2,k}, \ldots, \hat{Y}_{n,k}\}$, where $\hat{Y}_{i,k}$ denotes the generated state of vehicle $i \in [1, 2, \ldots, n]$ at time step $k$.

In summary, the traffic flow dynamics modeling problem is articulated as follows: Given the initial state $X_{0}$ of all following vehicles and the acceleration sequence $\mathbf{u} _{0:K}$ of the leading vehicle, the primary objective is to establish a model denoted as $\mathbf{H}(\cdot)$ within the embedding space of the state. This model enable us to generate trajectories for all following vehicles, denoted as $\hat{Y} _{1:K}$, in a manner that closely approximates the actual ground truth trajectory $Y_{1:K}$:

\begin{equation}
\label{Eq.Model.H}
\hat{Y} _{1:K} = \mathbf{H}(X_{0},\mathbf{u} _{0:K})
\end{equation}

\subsection{Method Framework\label{Sec.2.2}}

\begin{figure*}[!t]
\centering
\includegraphics[width=5.3in]{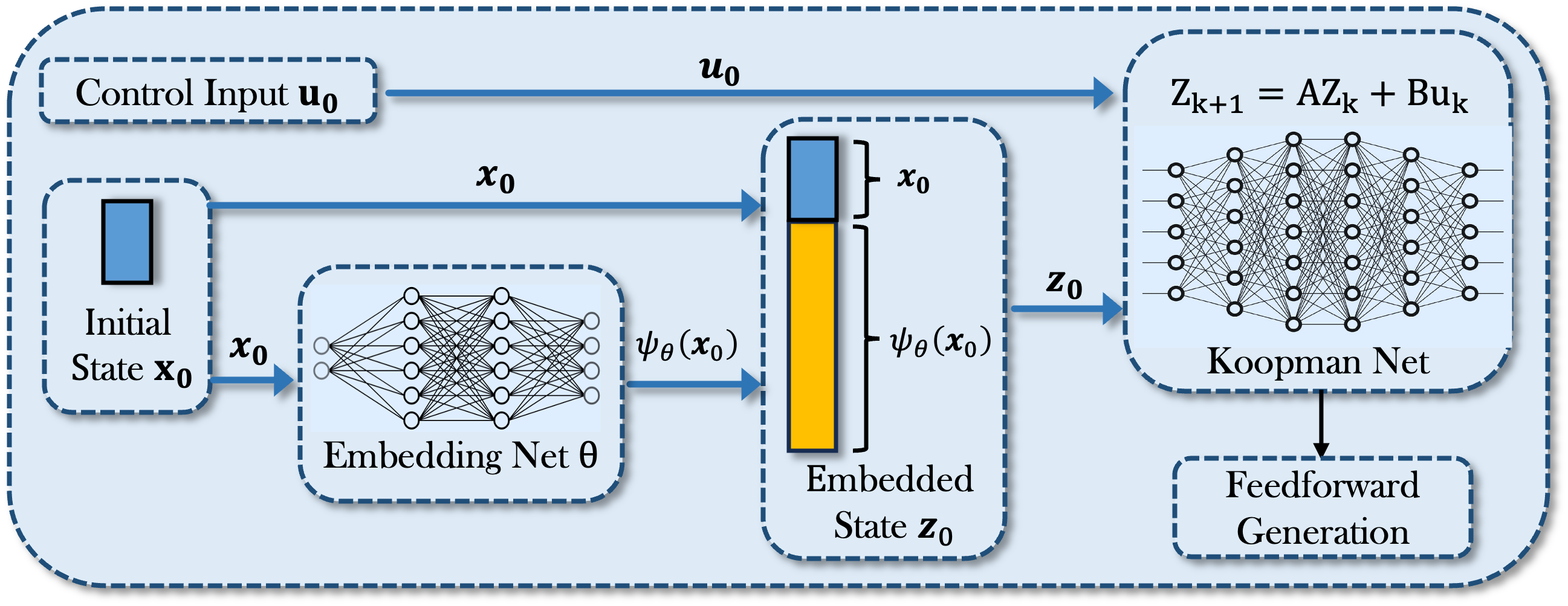}
\caption{Framework of the proposed methodology.}
\label{Fig.MethodFramework}
\end{figure*}

The proposed methodology framework is illustrated in Figure \ref{Fig.MethodFramework}. For modeling the traffic flow dynamics within the platoon, our initial step involves deriving a linear representation grounded in the theory of Koopman operators. This representation is expressed by the equation:

\begin{equation}
\label{Eq.Linear.Formula.0}
\mathbf{Z}_{k+1} = A\mathbf{Z}_{k} + B\mathbf{u}_{k}
\end{equation}

Following establishing the linear representation, we employ a deep learning framework consisting of two integral components: an Embedding Net and a Koopman Net. This framework concurrently learns both the embedding function and the Koopman operator. The Embedding Net utilizes a Deep Neural Network (DNN) \cite{liu2017survey} to train an embedding function, which is crucial in elevating the original state space of the traffic flow dynamics into the designated embedding space $\Psi$.  

Once the embedded space $\Psi$ is produced, we concatenate the original state $\mathbf{x}$ and the embedded state $\psi(\mathbf{X})$ to form the final embedded state $\mathbf{Z}$, which serves as the input for the Koopman Net. Then, the Koopman Net constitutes a linear model responsible for parameterizing the matrices $A$ and $B$. These matrices play a role in facilitating the reproduction process of subsequent traffic flow dynamics. By combining the strengths of deep learning and Koopman operator theory, the proposed methodology provides an innovative approach to capture the complex dynamics of traffic flow within a platoon, effectively bridging the gap between data-driven precision and physical interpretability.

\subsection{Method Formulation \label{Sec.2.3}}

\subsubsection{Definition of State Variables, Control Input, and Output \label{Sec.2.3.1}}

In this study, we define the state $\mathbf{X}_{k}$ for traffic flow dynamics using three fundamental variables integral to conventional car-following models:

\begin{itemize}
    \item \textbf{Spacing} $\mathbf{s}_{i,k}$, $i \in [1,2,\cdots,n]$: This variable quantifies the distance between vehicle $i$ and its preceding vehicle, denoted as vehicle $(i-1)$, at time $k$. 
    
    \item \textbf{Velocity} $\mathbf{v}_{i,k}$, $i \in [1,2,\cdots,n]$: This parameter represents the instantaneous speed of vehicle $i$ at time $k$.
    
    \item \textbf{Speed Difference} $\Delta \mathbf{v}_{i,k}$, $i \in [1,2,\cdots,n]$: This metric quantifies the difference in speed between vehicle $i$ and its preceding vehicle $(i-1)$, at time $k$.
\end{itemize}

Concerning the control input $\mathbf{u}_{k}$, we adopt the acceleration of the leading vehicle $\mathbf{a}_{0,k}$. The output, denoted as $\hat{Y}$, represents the generated trajectories of the following vehicles, specifically detailing the positions of following vehicles.

These variables are integral to most car-following models, incorporating the intricacies of driver-vehicle interactions in traffic scenarios. In summary, the states are defined as $\mathbf{X}_{k} = \begin{bmatrix}
\mathbf{s}_{i,k},
\mathbf{v}_{i,k},
\Delta \mathbf{v}_{i,k}
\end{bmatrix}^T_{i \in [1,2,\cdots,n]}$, and control inputs are defined as $\mathbf{u}_{k} = \mathbf{a}_{0,k}$.

\subsubsection{Fundamentals of Koopman Theory in Traffic Flow Dynamics Modeling \label{Sec.2.3.2}}

\begin{figure}[!t]
\centering
\includegraphics[width=3.5in]{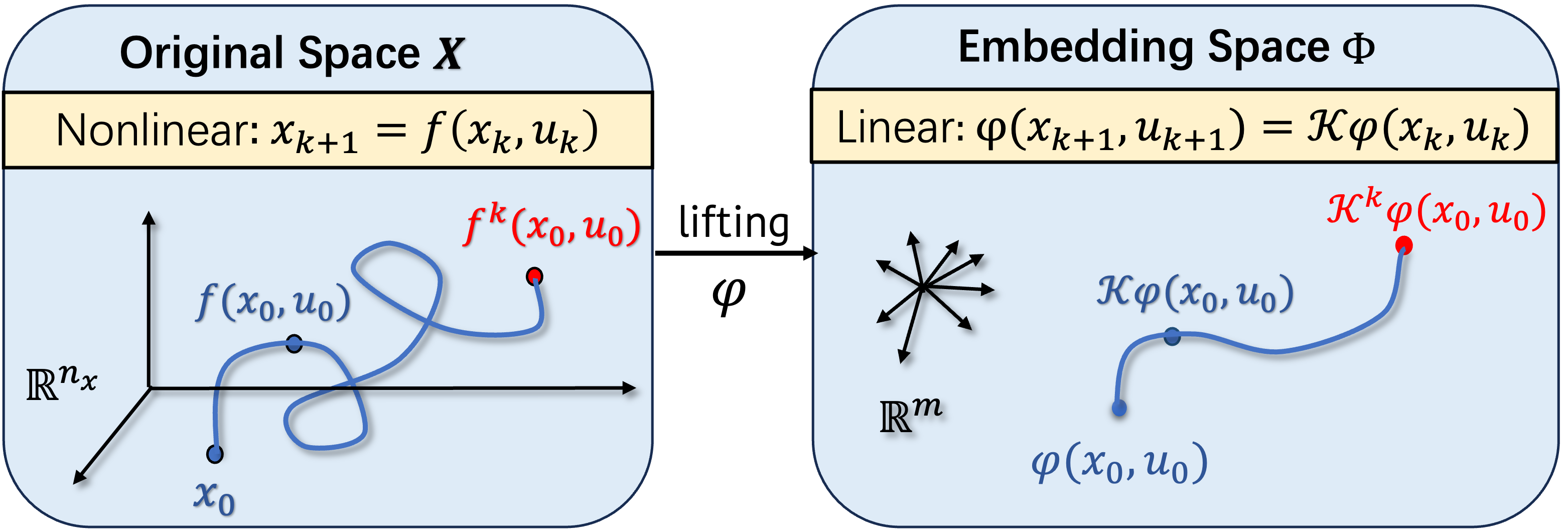}
\caption{Visual depiction of the lifting process transferring from original state space to embedding state space.}
\label{Fig.Lifting}
\end{figure}

Based on the model formulation, we delve into the principles of the Koopman theory. We consider discrete-time traffic flow dynamics governed by the nonlinear transformation $f: \mathbb{X} \to \mathbb{X}$:

\begin{equation}
\mathbf{X}_{k+1} = f (\mathbf{X} _{k}, \mathbf{u} _{k})
\label{Eq.NonlinearFunction}
\end{equation}
Here, $\mathbf{X} \in \mathbb{R}^{n_x}$ and $\mathbf{u} \in \mathbb{R}^{n_u}$ represent the state and control input of the traffic flow dynamics, respectively. Function $f$ maps the state space $\mathbb{X}$ onto itself, describing how the system's state evolves at each discrete time step $k$. 

To facilitate our analysis, we employ an embedding function $\psi: \mathbb{X} \to \Psi$, which transforms the original state of the traffic flow into an embedding state within the space $\Psi$. The Koopman operator unfolds the linear evolution within this embedding space:

\begin{equation}
\Psi = \{\psi(\mathbf{X}_1,\mathbf{u}_1), \psi(\mathbf{X}_2,\mathbf{u}_2), ... , \psi(\mathbf{X}_K,\mathbf{u}_K)\}
\end{equation}
Here, $\psi(\mathbf{X}_k,\mathbf{u}_k) \in \mathbb{R} ^ m$. The Koopman operator is a linear operator, which can be either finite or infinite dimensional, depending on the dynamics under consideration. Following the embedding, we apply the Koopman operator $\mathcal{K}: \Psi \to \Psi$ to the embedding space $\Psi$:

\begin{equation}
\mathcal{K} \psi \triangleq \psi \circ f
\label{Eq.KoopmanTheory}
\end{equation}

Upon applying the Koopman operator, it becomes evident that linearity prevails within the embedding space, as demonstrated by the equation:

\begin{equation}
\mathcal{K}(c_1 \psi_1 + c_2 \psi_2) = c_1 \mathcal{K} \psi_1 + c_2 \mathcal{K} \psi_2
\label{eq:KoopmanLinearity}
\end{equation}
where $\psi_1$ and $\psi_2$ represent two distinct embedding functions, while $c_1$ and $c_2$ denote scalar coefficients. Building on the definition in Equation (\ref{Eq.KoopmanTheory}), the dynamics of traffic flow within the embedding space $\Psi$ can be expressed as:

\begin{equation}
\mathcal{K} \psi(\mathbf{X} _{k}, \mathbf{u} _k) = \psi(f (\mathbf{X} _k, \mathbf{u} _k),\mathbf{u} _{k+1}) = \psi(\mathbf{X} _{k+1}, \mathbf{u} _{k+1})
\label{eq:EmbeddingFlow1}
\end{equation}

Alternatively, this relationship can be expressed as:

\begin{equation}
\psi(\mathbf{X} _{k+1}, \mathbf{u} _{k+1}) = \mathcal{K} \psi(\mathbf{X} _{k}, \mathbf{u} _k)
\label{eq:EmbeddingFlow2}
\end{equation}

The transformation of the state space, resulting from the application of the embedding function and Koopman operator, is visually represented in Figure \ref{Fig.Lifting}.

To disentangle the dynamics of the state $\mathbf{x}$ from the control input $\mathbf{u}$, we perform a partition of the embedding function $\psi$. This partitioning yields two distinct components, namely $\psi_{X}(\mathbf{X},\mathbf{u})$ and $\psi_{u}(\mathbf{X},\mathbf{u})$. The partitioned form of the equation is expressed as follows:

\begin{equation}
\begin{bmatrix}
    \psi_X(\mathbf{X}_{k+1}) \\
    \psi_u(\mathbf{u}_{k+1}) 
\end{bmatrix}
= 
\begin{bmatrix}
    \mathcal{K}_{11} & \mathcal{K}_{12} \\
    \mathcal{K}_{21} & \mathcal{K}_{22}
\end{bmatrix}
\begin{bmatrix}
    \psi_X(\mathbf{X}_{k}) \\
    \psi_u(\mathbf{u}_{k})
\end{bmatrix}
\label{Eq.PartitionXU1}
\end{equation}

Under the assumption that $\psi_{X}(\mathbf{X},\mathbf{u})$ depends on the state $\mathbf{X}$ and $\psi_{u}(\mathbf{X},\mathbf{u})$ is equivalent to the control input $\mathbf{u}$, we derive $\psi_{X}(\mathbf{X},\mathbf{u}) = \psi_{X}(\mathbf{X})$ and $\psi_{u}(\mathbf{X},\mathbf{u}) = \mathbf{u}$. Consequently, the evolution of traffic flow dynamics is represented as:
\begin{equation}
\begin{bmatrix}
    \psi_X(\mathbf{X}_{k+1}) \\
    \mathbf{u}_{k+1} 
\end{bmatrix}
= 
\begin{bmatrix}
    \mathcal{K}_{11} & \mathcal{K}_{12} \\
    \mathcal{K}_{21} & \mathcal{K}_{22}
\end{bmatrix}
\begin{bmatrix}
    \psi_X(\mathbf{X}_{k}) \\
    \mathbf{u}_{k} 
\end{bmatrix}
\label{Eq.PartitionXU2}
\end{equation}
We consider $\mathbf{u}$ as an external input. In accordance with (\ref{Eq.PartitionXU2}), a linear representation of the traffic flow dynamics can be expressed as:

\begin{equation}
    \psi_{X}(\mathbf{X} _{k+1}) = \mathcal{K}_{11}\psi_{X}(\mathbf{X} _{k})+\mathcal{K}_{12}\mathbf{u} _k
\label{Eq.LinearExpression0}
\end{equation}

\subsubsection{Koopman-Based Deep Learning Framework\label{Sec.2.3.3}}

\begin{figure}[!t]
\centering
\includegraphics[width=3.5in]{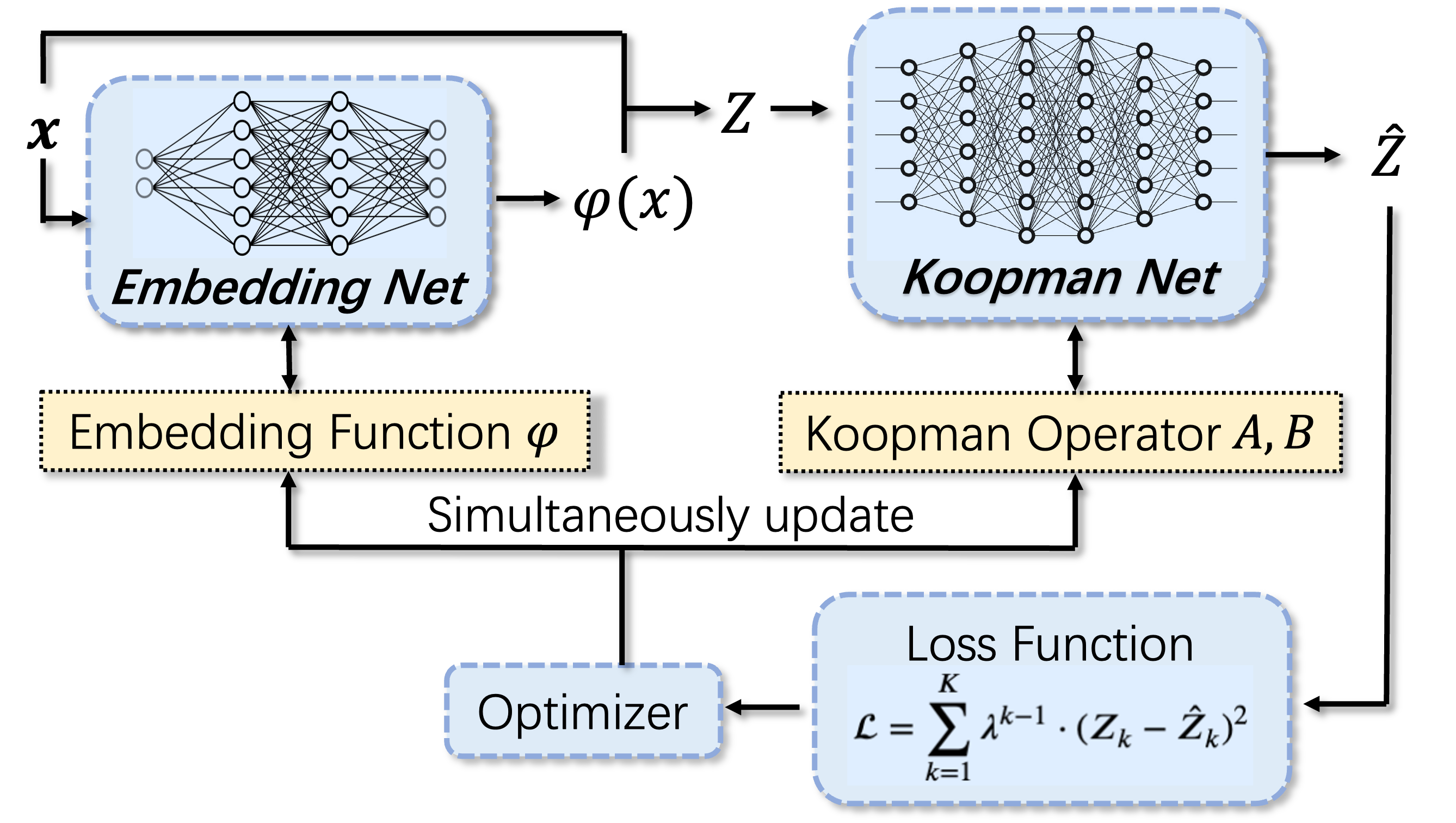}%
\caption{Training process of the proposed framework.}
\label{Fig.Training}
\end{figure}

The proposed methodology aims to capture the nonlinear traffic flow dynamics and represent it in a linear form as outlined in Equation (\ref{Eq.LinearExpression0}). To achieve this, we employ a deep learning framework that simultaneously trains the embedding function $\psi_X$ and the Koopman operator $\mathcal{K}$.

For the parameterization of the embedding function $\psi_{X}$, we utilize a DNN denoted as $\theta$, referred to as the \textit{Embedding Net}. The relationship between $\psi_{x}$ and the DNN can be expressed as:

\begin{equation}
\psi_{\theta}(\mathbf{X}_{k}) = \theta(\mathbf{X}_{k}), \quad \psi_{\theta}(\mathbf{X}_{k}) \in \mathbb{R}^{d}
\end{equation}
Here, $\psi_{\theta}: \mathbb{R}^{n_x} \to \mathbb{R}^{m}$ defines the parameterized \textit{Embedding Net}. To facilitate subsequent traffic flow reproduction processes, we require the original state to generate trajectories while maintaining simplicity in the \textit{Embedding Net} without needing an additional decoder. To achieve this, we concatenate the original state $\mathbf{X}_{k}$ with the parameterized state $\psi_{\theta}(\mathbf{X}_{k})$ to yield the final embedded state $\mathbf{Z}_{k}$ used for training:
\begin{equation}
\mathbf{Z}_{k} = \psi_{X}(\mathbf{X}_{k}) =
\begin{bmatrix}
    \mathbf{X}_{k} \\
    \psi_{\theta}(\mathbf{X}_{k})
\end{bmatrix}
, \quad \mathbf{Z}_{k} \in \mathbb{R}^{m}
\label{eq:ConcatenateEquation}
\end{equation}
Recognizing the structure of the embedded state \( \mathbf{Z} \) and its intricate relationship with the original state in dynamical systems, we developed a transformation to smoothly transition between these state spaces. This transformation is defined as:
\begin{equation}
\mathbf{X}_{k} = \mathbf{M} \mathbf{Z}_{k}
\label{Eq.EmbeddingToOriginal}
\end{equation}
The transformation relies on the matrix \(\mathbf{M}\), which features a customized block structure designed for this purpose:
\begin{equation}
\mathbf{M} = \begin{bmatrix}
I_{n_x} & 0 \\
\end{bmatrix}, \quad \mathbf{M} \in \mathbb{R}^{n_x \times (n_x+m)}
\end{equation}
Using the identity matrix \( I_{m} \) ensures the preservation of the original state components, with \(m\) representing the dimension of the embedded state. The zero matrix segment accommodates the additional embedded elements.

Thus, utilizing the \textit{Embedding Net}, we can express the evolution of the embedded state over time as follows:
\begin{equation}
\mathbf{Z}_{k+1} = \mathcal{K}_{11} \mathbf{Z}_{k} + \mathcal{K}_{12} \mathbf{u}_{k}
\end{equation}
Here, we define \(\mathbf{A} = \mathcal{K}_{11}\) and \(\mathbf{B} = \mathcal{K}_{12}\). To determine the Koopman operator matrices \(\mathbf{A}\) and \(\mathbf{B}\), we employ the \textit{Koopman Net}, a specialized linear network designed to accurately capture the system's underlying linear dynamics. This network facilitates the precise extraction and representation of the matrices responsible for governing the state transformations in our model.

Based on Equation (\ref{Eq.EmbeddingToOriginal}), with the given initial state \(\mathbf{X}_{0}\), we can iteratively derive the spacing data $\mathbf{\hat{X}}$ of the generated trajectory, following the sequential formulations:
\begin{equation}
\mathbf{Z}_{k+1} = \mathbf{A} \mathbf{Z}_{k} + \mathbf{B} \mathbf{u}_{k}
\label{eq: LinearExp1}
\end{equation}
\begin{equation}
\mathbf{\hat{X}}_{k+1} = \mathbf{M} \mathbf{Z}_{k+1}
\end{equation}
With the spacing data $\mathbf{\hat{X}}$ and the position of the leading vehicle, the generated trajectory $\hat{Y}$ can be calculated directly.

The training process is depicted in Figure \ref{Fig.Training}. We employ a loss function during the training phase based on exponentially decaying weighted squared error. While the Mean Squared Error (MSE)\cite{sayed2022inference} is a conventional metric for evaluating prediction accuracy, it treats all prediction errors uniformly without considering their temporal context. To overcome this limitation, we introduce a weighted loss function as a critical element of our network training strategy:
\begin{equation}
\mathcal{L} = \sum_{k=1}^{K} \lambda ^{k-1} \cdot (Z_k - \hat{Z}_k)^2
\end{equation}
Here, $\lambda$ represents the weight decay hyper-parameter. This weighting scheme enhances the network's capacity to capture intricate temporal dependencies and enhances its performance in tasks where accurate predictions across various time horizons are essential.

\begin{figure}[!t]
\centering
\includegraphics[width=3.5in]{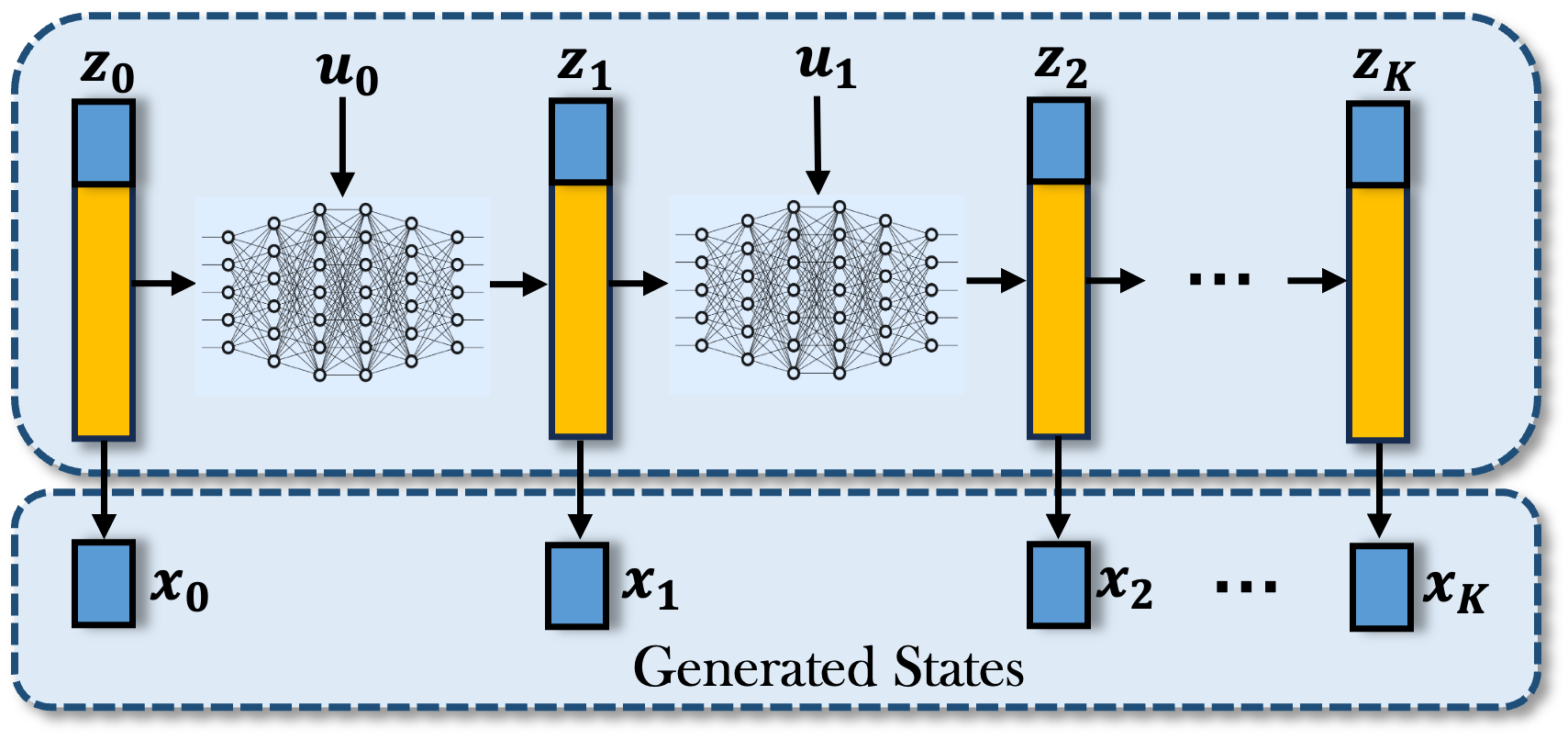}
\caption{The K-step feedforward trajectory generation process.}
\label{Fig.Feedforward}
\end{figure}

After training the \textit{Koopman Net}, we can accurately replicate the traffic flow dynamics via a feedforward trajectory generation process, as illustrated in Figure \ref{Fig.Feedforward}.

\section{Experiment results and analysis \label{Sec.3}}

\subsection{Data Preparation \label{Sec.3.1}}

To showcase the broader applicability of the proposed method, we apply the proposed methodology to a real-world scenario using the Next Generation Simulation (NGSIM) dataset \cite{punzo2011assessment}. Specifically, we extract data pertaining to a platoon of six vehicles, essentially $n=6$, from the NGSIM dataset.
Each individual car-following scenario in our analysis spans 350 time steps, equivalent to 35 seconds, with a temporal interval of 0.1 seconds separating consecutive time steps. To facilitate the subsequent training process, we divide our dataset into two distinct subsets: a training set and a test set, divided in a 4:1 ratio, splitting a total of 50 trajectories.
\subsection{Performance Comparison among Traffic Dynamics Modeling Approaches\label{Sec.3.2}}
To show the improved performance of our method in modeling traffic flow dynamics, we have selected three prominent car-following analysis approaches to serve as baseline models (i.e., IDM \cite{kesting2010enhanced}, DMD\cite{kutz2016dynamic}, LSTM \cite{huang2018car}). These approaches encompass both model-based and data-driven methodologies. We employ two fundamental evaluation metrics, specifically the Root Mean Square Error (RMSE) and the Mean Absolute Error (MAE)\cite{hodson2022root} of positions, to assess the performance of our traffic flow dynamics modeling. 

\begin{table}[!t]
\caption{Generation Error Obtained by Different Models \label{Tab.Error}}
\centering
\begin{tabular}{|c||c|c|} 
\hline
{} & RMSE (m) & MAE (m) \\ 
\hline
Deep Koopman & \textbf{2.6806} & \textbf{1.9540} \\
LSTM-based & 3.2783 & 2.3546\\
IDM & 5.1757 & 3.4169\\
DMD & 13.1078 & 4.4453\\
\hline
\end{tabular}
\end{table}

Based on the evaluation metrics, we conduct a comprehensive analysis to assess the effectiveness of all baseline models. As illustrated in Table \ref{Tab.Error}, our proposed method exhibits superior performance compared to all baseline models. It is characterized by achieving the lowest RMSE at 2.6806 meters and the lowest MAE at 1.9540 meters. The second-lowest generation error is observed in the LSTM-based model, which, although commendable, records a higher RMSE of 3.2783 meters and an MAE of 2.3546 meters. In contrast, the IDM and DMD exhibits highest generation errors in both RMSE and MAE, thus representing the weakest performance among the baseline models. Comparatively, DMD performs worst, suggesting the traffic dynamic system is nonlinear.

\begin{figure*}[!t]
\centering
\includegraphics[width=5.5in]{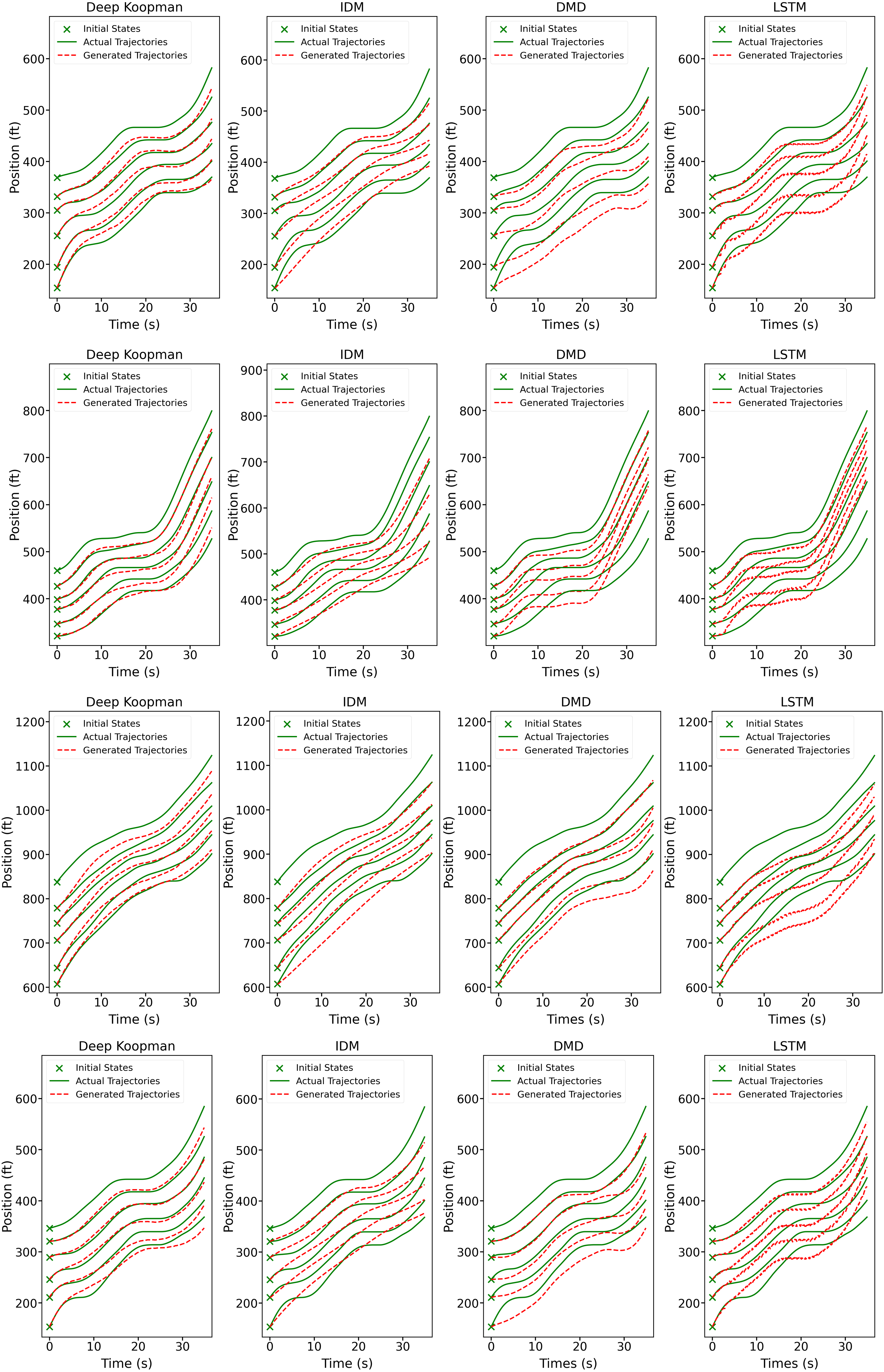}%
\caption{An reproduced trajectory Example among all baseline models.}
\label{Fig.Trajectory}
\end{figure*}

Compared to the LSTM-based model, a benchmark for optimized AI-based models, our method demonstrates a systematic improvement of approximately $15\%$ in RMSE. This improvement underscores our method's superior ability to learn the nonlinear dynamics of traffic flow with enhanced precision compared to other data-driven techniques. Furthermore, a linear state space representation of the traffic flow dynamics in the embedding space is derived. This representation describes how the current state, influenced by the system's inherent dynamics and external inputs, evolves into the future state. 
To provide a comprehensive assessment of the effectiveness of our approach in modeling traffic dynamics and its ability to reproduce traffic flow, we conducted a series of experiments aimed at generating the trajectories of a platoon. This was accomplished using the real-world data described in Section \ref{Sec.3.1}. The primary goal of this visualization (see Figure \ref{Fig.Trajectory}) is to illustrate the performance of traffic dynamics modeling.

In this experiment, we evaluated our approach alongside baseline models, including the LSTM-based model, IDM, and DMD. The task involved reproducing the platoon's traffic flow dynamics over a 350-time step period, using only the initial state and control variables as input. 

Upon closer examination of the generated trajectories, it becomes evident that our proposed method consistently maintains accuracy over an extended duration. Our model excels in capturing both the short-term and long-term dynamics of the traffic flow. The trajectories it generates closely follow the ground truth (solid green line) with remarkable precision. In contrast, when we compare the performance of our approach with that of the baseline models, we observe that the latter exhibits initial accuracy but gradually accumulates significant errors as time and vehicle number progresses. This is particularly evident in the dashed red trajectories representing the output of the baseline models. Notably, some approaches, particularly DMD, struggle to produce stable trajectories across all scenarios, resulting in erratic and less reliable predictions. The visual observations are in strong agreement with the quantitative findings presented in Table 1, where our method consistently outperforms the baseline models across various metrics. 

\subsection{Phase Plane Analysis \label{Sec.3.3}}

\begin{figure*}[!t]
\centering
\includegraphics[width=7in]{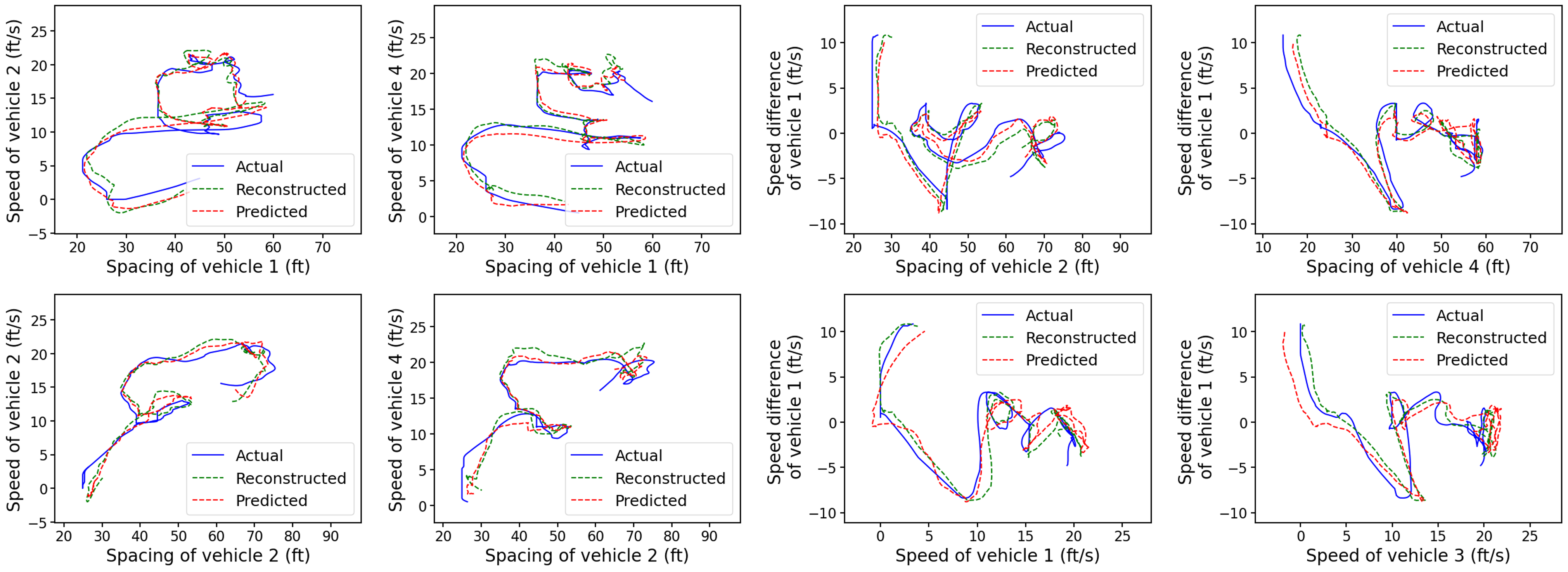}%
\caption{Phase Plane Analysis}
\label{Fig.PhasePlane}
\end{figure*}

We further conduct a comparative analysis using phase planes, which provide a geometric representation of the orbits of traffic flow dynamics. Three types of phase planes are generated based on different pairs of state variables: spacing vs. speed, spacing vs. speed difference, and speed vs. speed difference. In each phase plane, we compare the motion trajectories between the ground truth dynamics, the predicted trajectories, and the reconstructed trajectories obtained by our proposed approach. These phase planes reveal intricate and nonlinear patterns crucial for understanding the dynamic behavior of traffic flow. By comparing the true dynamics with those reproduced by our approach, the proposed approach's ability to model nonlinear traffic flow dynamics is assessed.

The phase planes, depicted in Figure \ref{Fig.PhasePlane}, collectively offers a comprehensive visualization of the complex nonlinear dynamics of vehicular platoons.  This reconstruction employs the Koopman operator $A$ and $B$ through the formula $\mathbf{Z}_{k+1} = \mathbf{A} \mathbf{Z}_{k} + \mathbf{B} \mathbf{u}_{k}$, where $\mathbf{Z}_{k}$ is the ground truth state at time step $k$. For the predicted phase portraits, they are constructed using the well-trained feedforward network for ten time steps. In this process, we predict $\mathbf{Z}_{k+10}$ using the ground truth state at time step $k$. All three sets of trajectories closely align with the ground truth, demonstrating that our proposed method excels in comprehending the intricate relationships among different state variables. This showcases its capacity to capture and reproduce the nonlinear dynamics of traffic flow with precision and accuracy.

\subsection{Stability Analysis of Traffic Flow \label{Sec.3.4}}

Our approach not only models the nonlinear traffic flow dynamics with high accuracy but also provides a physical linear representation of traffic flow. This representation enables the application of several well-established analysis tools typically reserved for linear models. One such tool is the stability analysis of traffic flow dynamics.

The stability issue holds paramount significance within the domain of traffic dynamics as it pertains to deciphering the propagation of disturbances and mitigating the potential for traffic fluctuations and congestion. Given its profound implications for safety, dependability, and operational efficiency, the subject of traffic stability has undergone extensive examination. In this section, we employ both local stability and string stability analysis techniques to investigate the traffic flow dynamics. Traditionally, these two stability analyses are exclusively applicable to linear models. Within our approach, the inherently complex nonlinear traffic flow dynamics become amenable to physical analysis, allowing us to gain valuable insights into their stability characteristics.

\subsubsection{Local Stability \label{Sec.3.4.1}}

\begin{figure}[!t]
\centering
\includegraphics[width=2.5in]{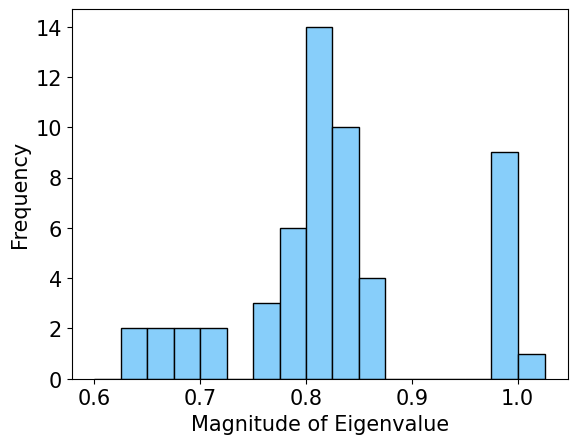}
\caption{Local Stability Analysis}
\label{Fig.LocalStability}
\end{figure}

Local stability, also known as internal stability, pertains to a vehicle's ability to resolve disturbances (deviations from equilibrium spacing, speed difference, or acceleration) within its local vicinity \cite{zhou2020stabilizing}. 

We adopt the definition of internal stability from \cite{chen1984linear}. In the context of a zero-input Linear Time-Invariant (LTI) system, internal stability implies that the system is either asymptotically or marginally stable.

\begin{defn}
    For a discrete zero-input LTI system defined as:
\begin{equation}
   \mathbf{X}(t+1) = \mathbf{A}\mathbf{X}(t)    \text{, \space \space \space} \mathbf{X}(0) = \mathbf{X}_0
    \label{eq:LTI}
\end{equation}
the system is considered asymptotically stable if $\mathbf{x}(t) \to 0$ as $t \to \infty$ for every initial condition $\mathbf{x}_0$,
\end{defn}

\begin{defn}
    For a zero-input LTI system defined as:
\begin{equation}
    \mathbf{X}(t+1) = \mathbf{A}\mathbf{X}(t)    \text{, \space \space \space} \mathbf{X}(0) = \mathbf{X}_0
    \label{eq:LTI}
\end{equation}
the system is considered marginally stable if $\mathbf{X}(t)$ remains bounded as $t \to \infty$ for every $\mathbf{X}_0$.
\end{defn}

In the context of discrete nonlinear traffic flow dynamics, as described in section \ref{Sec.2.3.2}, the linear representation utilizing the Koopman operator is defined as follows:
\begin{equation}
    \mathbf{Z}_{k+1} = \mathbf{A} \mathbf{Z}_{k} + \mathbf{B} \mathbf{u}_{k}
    \label{eq:LinearRepre2}
\end{equation}
Since stability is an intrinsic characteristic of the system, it depends solely on matrix $\mathbf{A}$ \cite{bishop2011modern}. Consequently, we assume that the control input $\mathbf{u}_k$ is held at zero, leading to the simplified equation:
\begin{equation}
    \mathbf{Z}_{k+1} = \mathbf{A} \mathbf{Z}_{k}
    \label{eq:SimplifiedLinearExp}
\end{equation}

The local stability of the discrete platoon system can be ascertained by scrutinizing the eigenvalues of the Koopman operator $\mathbf{A}$. Specifically, the system is considered internally stable if the magnitude of every eigenvalue $\lambda_i$ of matrix $\mathbf{A}$ is strictly less than or equal to 1. Conversely, the system is deemed unstable if the magnitude of any eigenvalue is greater than or equal to 1.

A Koopman matrix $A$ representing the traffic flow dynamics as described in section \ref{Sec.3.1} is derived from the Koopman Net following the training process. The Koopman matrix $A$ is a square matrix with dimensions $55\times55$. Utilizing Equation (\ref{eq:SimplifiedLinearExp}) and the definition of local stability, we compute the eigenvalues $\lambda_i$ of the Koopman matrix $A$. There are 55 distinct eigenvalues of matrix A, which confirms its diagonalizability. The distribution of the magnitudes of the 55 eigenvalues of the Koopman matrix $A$ is illustrated in Figure \ref{Fig.LocalStability}. Notably, the maximum magnitude of an eigenvalue is strictly greater than 1, indicating the presence of at least one eigenvalue with a magnitude exceeding 1. Consequently, in accordance with the definition of local stability, platoons consisting of HDVs are deemed locally unstable.

This local stability analysis aligns with real-world conditions where the driving behavior of HDVs exhibits diversity, stochasticity, and lacks adherence to a specific car-following strategy. Some drivers may tend to overreact, while others may underreact, resulting in a cumulative effect that can lead to local instability. This consistency validates the physical analyzable feature of our approach.

\subsubsection{String Stability \label{Sec.3.4.2}}

String stability, a critical property of traffic flow, pertains to traffic's ability to dampen disturbances, such as the frequent acceleration and deceleration observed in leading vehicles. Current theoretical frameworks are primarily applicable to linear or linearized car-following (CF) laws. There are limited instances where nonlinear CF string stability analysis is conducted, mainly through describing function analysis, focusing on specific and relatively simple nonlinearities within CF laws, such as vehicular acceleration and deceleration boundaries. Complex nonlinearities are often beyond the scope of these analyses. However, the proposed Koopman-based approaches enable the use of data-driven techniques to enhance the precision of interpreting nonlinear traffic dynamics. In this specific context, we investigate the propagation of disturbances within the H2-norm function space \cite{naus2010string}. This function space is characterized by an "energy" norm, which is elaborated upon below. By the definition of head-to-tail string stability, a vehicular string is head-to-tail string stable if and only if the following condition holds for a platoon with size $N$ \cite{zhou2020stabilizing}:
\begin{equation}
    \frac{||a_{N}(s)||_2}{||a_{0}(s)||_2} \leq 1
    \label{eq:string_stability_condition}
\end{equation}
Here, $a_{0}(s)$ represents the acceleration of the leading vehicle, vehicle $0$, in the frequency domain; $a_{N}(s)$ represents the acceleration of the last vehicle in the platoon, where $N=5$ based on the setting of our experiment. The variable $s$ is defined as:$s = j\omega $, where $\omega > 0$ represents the frequency, and $j$ is the imaginary unit. 
By applying the Cauchy inequality \cite{zhou2020stabilizing}, we can derive the following relationship:
\begin{equation}
    \frac{||a_{N}(s)||_2}{||a_{0}(s)||_2} \leq ||G(s)||_\infty = \sup|G(j\omega)|, \forall \omega > 0
\end{equation}
Here, $G(s)$ is the transfer function describing disturbance propagation through the vehicular string in the frequency domain.Consequently, it becomes evident that ensuring $sup|G(j\omega)| \leq 1$ for all $\omega > 0$ is sufficient to establish string stability.

\begin{figure}[!t]
    \centering
    \includegraphics[width=2.5in]{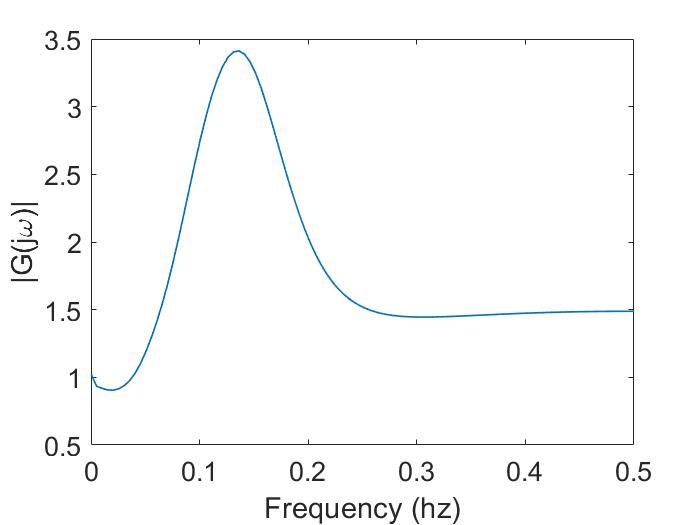}
    \caption{String Stability Analysis}
    \label{Fig.StringStability}
\end{figure}

The traffic flow dynamics we are examining are discrete, represented by the state space function according to Equation (\ref{eq: LinearExp1}):
\begin{equation}
    \mathbf{Z}_{k+1} = \mathbf{A} \mathbf{Z}_{k} + \mathbf{B} \mathbf{u}_{k}
    \label{eq:LinearExp3}
\end{equation}
Here, at time step $k$, the control input $\mathbf{u}_{k}$ corresponds to the acceleration of vehicle 0:
\begin{equation}
    \mathbf{u}_{k} = \mathbf{a}_{0,k}
\end{equation}

To analyze these dynamics in continuous time, we apply the \textit{zero-hold order} method to transform Equation (\ref{eq:LinearExp3}) into a continuous-time state-space function. This transformation is performed using the \textit{d2c()} function in MATLAB, resulting in the following continuous-time representation:
\begin{equation}
    \mathbf{\Dot{Z}}(t) = \mathbf{A}_{c} \mathbf{Z}(t) + \mathbf{B}_{c} \mathbf{u}(t)
    \label{eq:ContEq}
\end{equation}
In this equation, $\mathbf{A}_{c}$ denotes the Koopman operator $\mathbf{A}$ within the continuous-time domain, and likewise, $\mathbf{B}_{c}$ represents the Koopman operator $\mathbf{B}$ in continuous time.

To extract the velocity of vehicle $i$ from the state vector $\mathbf{Z}(t)$, we define $C_{i}$ as $\mathbf{e}_{i+5}$, where $i \in \{1,2,3,4,5\}$. The velocity of vehicle $i$ is then calculated as:
\begin{equation}
    v_{i}(t) = C_{i} \cdot \mathbf{Z}(t), \quad 1 \leq i \leq 5
    \label{eq:vel}
\end{equation}

To analyze the string stability, we apply Laplace transforms to the continuous-time Equation (\ref{eq:ContEq}) and Equation (\ref{eq:vel}). This transformation yields the following relationship:

\begin{equation}
    C_i(sI-\mathbf{A}_c)^{-1} \cdot \mathbf{B}_c = \frac{\mathbf{v}_i(s)}{\mathbf{a}_0(s)}
    \label{eq:L_T_relation}
\end{equation}
In this context, $\mathbf{v}_i(s)$ represents the velocity of vehicle $i$ in the frequency domain, and $\mathbf{a}_0(s)$ represents the acceleration of vehicle $i$ in the frequency domain $s$. Based on Equation (\ref{eq:L_T_relation}), we derive the transfer function between vehicle $5$ and the leading vehicle as follows:

\begin{equation}
    G(s) = \frac{\mathbf{a}_5(s)}{\mathbf{a}_0(s)} = s \cdot \frac{\mathbf{v}_5(s)}{\mathbf{a}_0(s)} = s \cdot C_5(sI-\mathbf{A}_c)^{-1} \cdot \mathbf{B}_c
\end{equation}

By substituting $s=j\omega$, we obtain $|G(j\omega)|$, which characterizes the damping ratio that describes the acceleration relationship between vehicle $5$ and the leading vehicle. Previous studies have demonstrated that $|G(j\omega)|$ being consistently less than 1 across the entire frequency range is a sufficient condition for $H_2$-norm string stability \cite{zhou2020stabilizing}.

Utilizing the Koopman operator matrices $A$ and $B$ obtained through extensive training of the Koopman Next, we calculate the transfer function magnitude value, $|G(j\omega)|$. The results are illustrated in Figure \ref{Fig.StringStability}. Upon examining the outcomes, a notable observation emerges. At a lower frequency range $0.00 Hz \leq s \leq 0.05 Hz$, the disturbance of is attenuated as it passes through the vehicular string; at a higher frequency range $s \geq 0.05 Hz$, the disturbance is amplified as it passes through the vehicular string; At frequency $s \approx 0.13 Hz$, $|G(j\omega)|$ reaches a maximum value near $3.5$, and when the frequency $s \geq 0.25 Hz$, $|G(j\omega)|$ tends to be stable around $1.5$. This finding generally implies that the HDV platoon exhibits characteristics of string instability, where disturbances can potentially amplify as they propagate through the vehicle string, during traffic oscillation known for higher-frequent ocillatory features.

\section{Conclusion \label{Sec.4}}

This research developed an AI-based Koopman approach to model nonlinear traffic dynamics, particularly under traffic oscillation, with physical interpretability and analyzability. We first introduce the Koopman Operator Theory and, based on this theory, conduct a linear representation of the nonlinear traffic flow dynamics in the embedding space. Utilizing the Koopman Operator Theory, a neural network-based framework was proposed to generate the embedding function that lifts the original space into the embedding space, along with the Koopman Operator in the linear representation. Notably, this methodology synergizes the physical interpretability and analyzability of model-based methods with the heightened precision inherent in data-driven approaches.

The performance of the proposed approach is examined through a series of simulation experiments utilizing the NGSIM dataset. Initially, we benchmark our method against baseline models, including IDM, LSTM, and DMD, in terms of its ability to reproduce nonlinear traffic flow dynamics. The results consistently demonstrate the superior performance of our approach, particularly in accurately reproducing traffic dynamics, even under the traffic oscillation. Furthermore, we employ phase plane plots to visually compare the ground truth traffic dynamics with those reproduced using our methodology, highlighting its exceptional capacity to model the complex nonlinear traffic behavior. Additionally, we conduct local stability and string stability analyses, underscoring the physical analyzability of our proposed method.

Nevertheless, future studies can be implemented based on the current methodology. For instance, it can be extended to encompass nonlinear traffic systems featuring Autonomous Vehicles. 

\section{References Section}

\bibliographystyle{IEEEtran}
\bibliography{IEEEabrv,ref}
\section{Biography Section}
\begin{IEEEbiography}[{\includegraphics[width=1in,height=1.25in,clip,keepaspectratio]{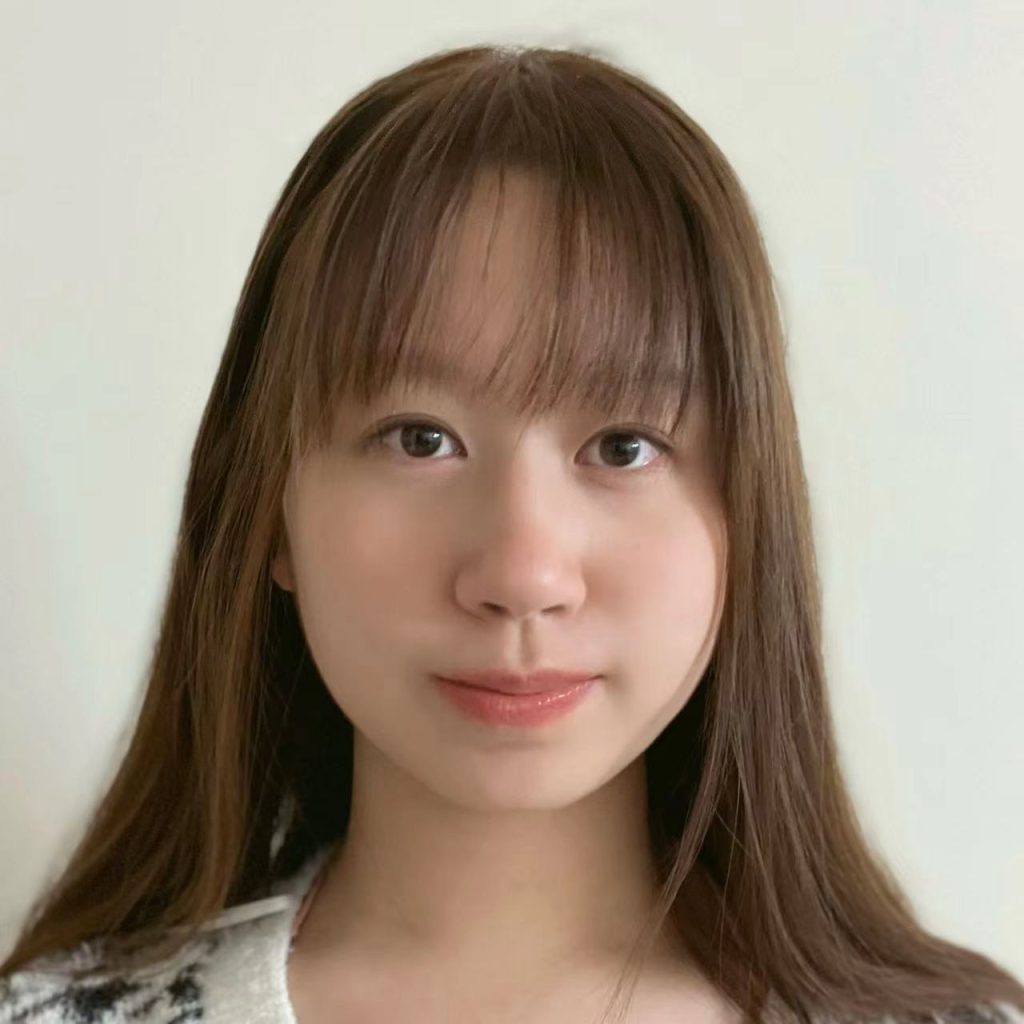}}]{Kexin Tian}
serves as a research assistant at Texas A\&M University. She received her M.S. degree in Civil and Environmental Engineering from University of Wisconsin Madison, WI, USA, in 2024, and the B.S degrees in Computer Science and Mathematics from University of Wisconsin Madison, in 2022. She is currently pursuing her Ph.D. degree in Civil and Environmental Engineering with Texas A\&M University, College Station, TX, USA. Her main research directions are traffic flow prediction, intelligent transportation systems, and machine learning.
\end{IEEEbiography}
\begin{IEEEbiography}[{\includegraphics[width=1in,height=1.25in,clip,keepaspectratio]{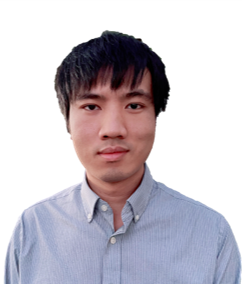}}]{Haotian Shi}
serves as a research associate at University of Wisconsin-Madison. He received his Ph.D. degree in Civil and Environmental Engineering from the University of Wisconsin-Madison in May 2023. He also received three M.S. degrees in Power and Machinery Engineering (Tianjin University, 2020), Civil and Environmental Engineering (UW-Madison, 2020), and Computer Sciences (UW-Madison, 2022). His main research directions are connected and automated vehicles, intelligent transportation systems, traffic crash data analysis, and deep reinforcement learning.
\end{IEEEbiography}
\begin{IEEEbiography}[{\includegraphics[width=1in,height=1.25in,clip,keepaspectratio]{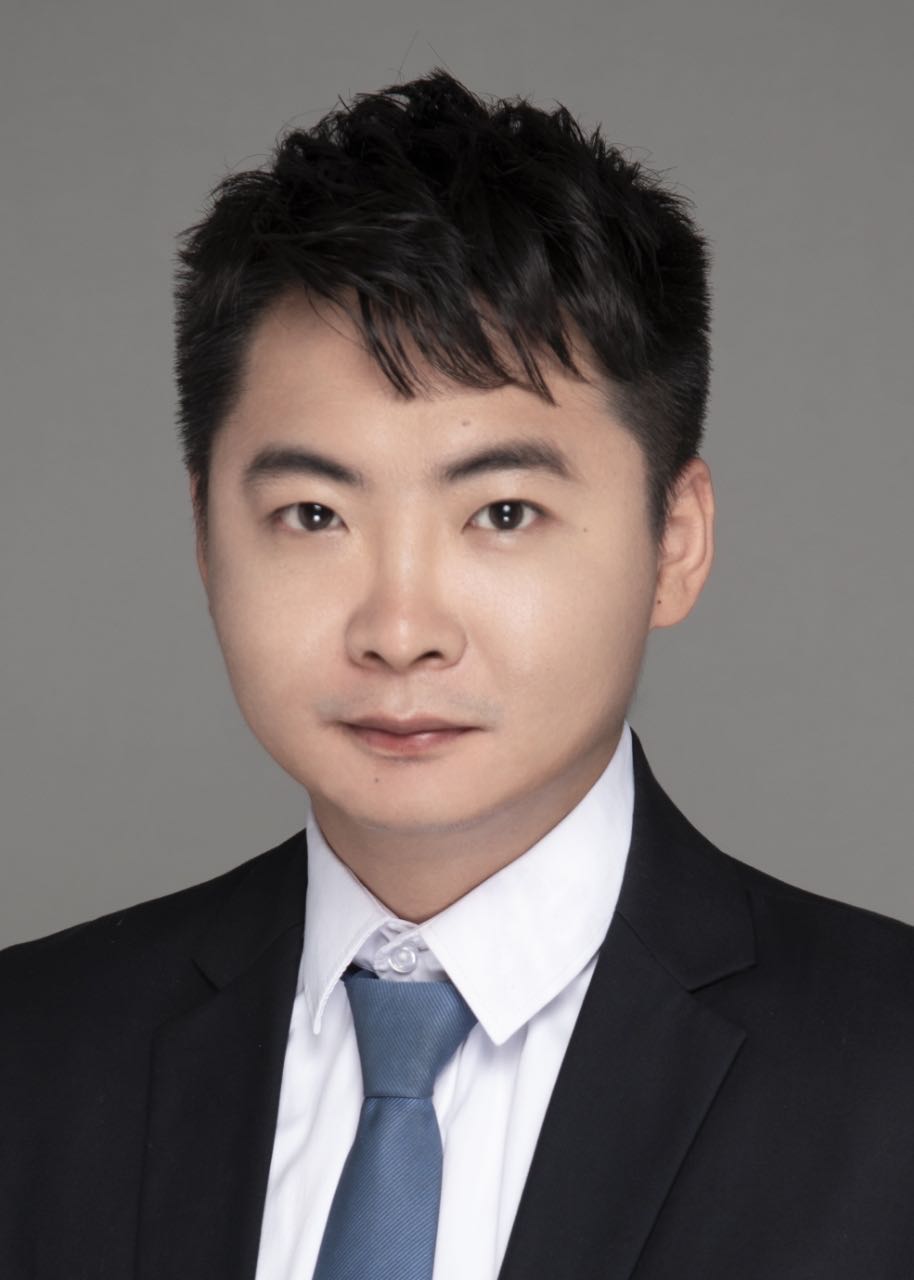}}]{Yang Zhou}
received the Ph.D. degree in Civil and Environmental Engineering from University of Wisconsin Madison, WI, USA, in 2019, and the M.S. degree in Civil and Environmental Engineering from  University of Illinois at Urbana-Champaign, Champaign, IL, USA, in 2015. He is currently an assistant professor in the Zachry Department of Civil and Environmental Engineering, and Career Initiation Fellow in the Institute of Data Science, Texas A\&M University. Before joining Texas A\&M, he is a postdoctoral researcher in civil engineering, University of Wisconsin Madison, WI, USA. He is current an member in TRB traffic flow theory CAV subcommittee, network modeling CAV subcommittee and American Society of Civil Engineering TDI-AI committee. His main research directions are connected automated vehicles robust control, interconnected system stability analysis, traffic big data analysis, and microscopic traffic flow modeling.
\end{IEEEbiography}
\begin{IEEEbiography}[{\includegraphics[width=1in,height=1.25in,clip,keepaspectratio]{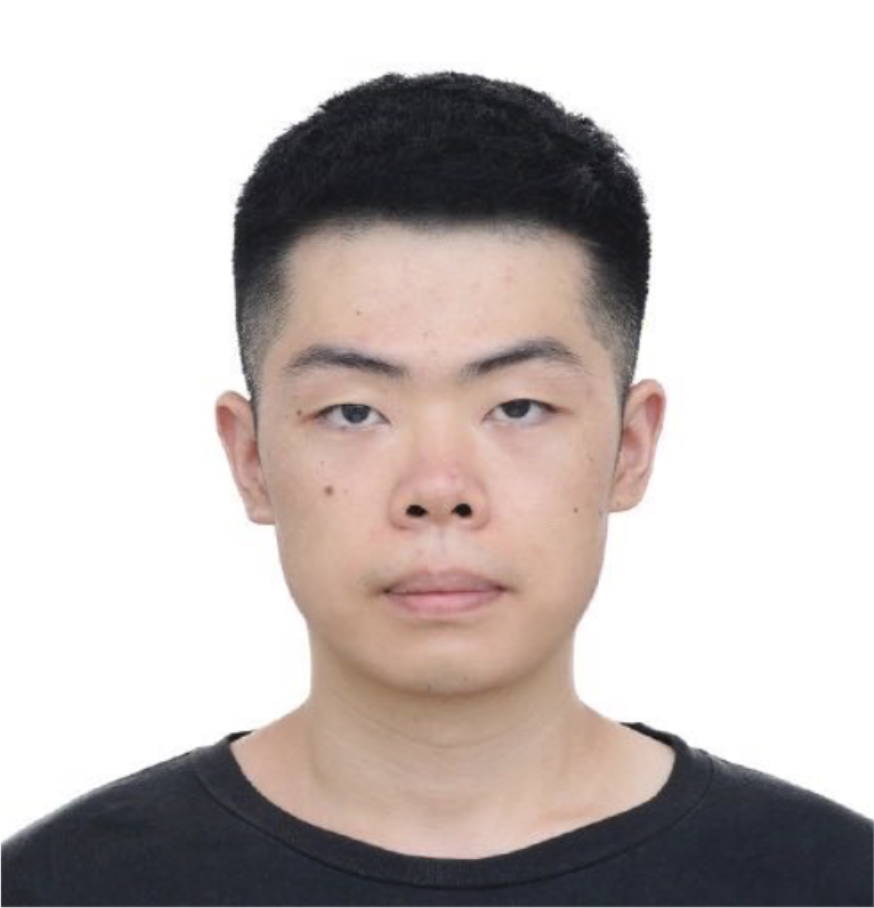}}]{Sixu Li}
received the B.S. degree in Engineering Mechanics from Hunan University, Changsha, China, in 2020, and the master’s degree in Mechanical Engineering from UC Berkeley, Berkeley, CA, USA, in 2022. He is currently pursuing the Ph.D. degree in Civil and Environmental Engineering with Texas A\&M University, College Station, TX, USA.
His current research interests include dynamics and control, MPC, and optimization for autonomous driving,  intelligent transportation systems, and robotics.
\end{IEEEbiography}

\vspace{11pt}

\vfill

\end{document}